\documentclass[lettersize,journal]{IEEEtran}
\usepackage{amsmath,amsfonts}
\usepackage{algorithmic}
\usepackage{algorithm}
\usepackage{array}
\usepackage[caption=false,font=normalsize,labelfont=sf,textfont=sf]{subfig}
\usepackage{textcomp}
\usepackage{stfloats}
\usepackage{url}
\usepackage{verbatim}
\usepackage{graphicx}
\usepackage{cite}
\usepackage[colorlinks, linkcolor=black, citecolor=blue, urlcolor=blue]{hyperref}
\usepackage{graphicx} % Required for inserting images
\usepackage{caption}
\usepackage{amsmath}
\usepackage{listings}
\usepackage{multirow}
\usepackage{caption}
\usepackage{subcaption}
\usepackage{tcolorbox}
\usepackage{hyperref}
\usepackage{url}
\usepackage{listings}
\usepackage{array}
\usepackage{booktabs}
\usepackage{color}

\usepackage{pifont}
\usepackage{amssymb}
\begin{document}

\title{MAGNET: A Multi-Graph Attentional Network for Code Clone Detection}

\author{Zixian~Zhang~and~Takfarinas~Saber%
\thanks{Zixian Zhang is with CRT-AI, University of Galway, Ireland (email: Z.Zhang15@universityofgalway.ie).}%
\thanks{Takfarinas Saber is with Lero, School of Computer Science, University of Galway, Ireland (email: takfarinas.saber@universityofgalway.ie).}%
\thanks{Corresponding author: Zixian Zhang.}%
}

% The paper headers
% \markboth{Journal of \LaTeX\ Class Files,~Vol.~14, No.~8, August~2021}%
% {Shell \MakeLowercase{\textit{et al.}}: A Sample Article Using IEEEtran.cls for IEEE Journals}

% \IEEEpubid{0000--0000/00\$00.00~\copyright~2021 IEEE}
% Remember, if you use this you must call \IEEEpubidadjcol in the second
% column for its text to clear the IEEEpubid mark.

\maketitle

\begin{abstract}
Code clone detection is a fundamental task in software engineering that underpins refactoring, debugging, plagiarism detection, and vulnerability analysis. Existing methods often rely on singular representations such as abstract syntax trees (ASTs), control flow graphs (CFGs), and data flow graphs (DFGs), which capture only partial aspects of code semantics. Hybrid approaches have emerged, but their fusion strategies are typically handcrafted and ineffective. In this study, we propose MAGNET, a multi-graph attentional framework that jointly leverages AST, CFG, and DFG representations to capture syntactic and semantic features of source code. MAGNET integrates residual graph neural networks with node-level self-attention to learn both local and long-range dependencies, introduces a gated cross-attention mechanism for fine-grained inter-graph interactions, and employs Set2Set pooling to fuse multi-graph embeddings into unified program-level representations. Extensive experiments on BigCloneBench and Google Code Jam demonstrate that MAGNET achieves state-of-the-art performance with an overall F1 score of 96.5\% and 99.2\% on the two datasets, respectively. Ablation studies confirm the critical contributions of multi-graph fusion and each attentional component. Our code is available at \url{https://github.com/ZixianReid/Multigraph_match}

\end{abstract}


\section{Introduction}
\label{sec:introduction}
Code clone detection is a fundamental task in software engineering, focused on identifying duplicated or highly similar code fragments within a software repository~\cite{saini2018code}. Such clones often arise from common development practices, including copy-pasting, reusing code templates, or implementing similar functionalities independently. The presence of clones can lead to software redundancy, propagate bugs, and decrease code stability, making their detection critical for improving software quality~\cite{roy2007survey}. As a foundational task, clone detection underpins a wide range of downstream software engineering activities, including refactoring~\cite{tsantalis2017clone, tsantalis2015assessing}, code search~\cite{nishi2018scalable}, debugging~\cite{li2006cp, ebrahimi2019hmm}, plagiarism detection~\cite{fokam2021influence, cheers2020novel}, and vulnerability analysis~\cite{xiao2017evolution, sun2021vdsimilar}.

A central challenge in clone detection lies in how code fragments are represented. Early approaches relied on handcrafted lexical and syntactic features, which proved effective for identifying Type-1 and Type-2 clones but struggled with structurally flexible or semantically similar clones (Type-3/4)~\cite{sajnani2016sourcerercc, bellon2007comparison}. To address these limitations, modern approaches increasingly extract latent embeddings from code fragments and compute distances between them to measure functional similarity~\cite{siow2022learning}. Depending on the representation approach, existing methods can be broadly categorized into token-based~\cite{sajnani2016sourcerercc, wang2018ccaligner, golubev2021multi}, tree-based~\cite{wei2017supervised, zhang2019novel, jiang2022hierarchical}, and graph-based~\cite{liu2023learning, wang2020detecting, mehrotra2023improving} methods, each with distinct limitations. Token-based methods treat source code as sequences of lexical tokens but neglect the hierarchical and structural relationships inherent within code. Tree-based methods, primarily utilizing abstract syntax trees (ASTs), effectively capture syntactic hierarchy but fail to explicitly model dynamic semantics such as control flow and data dependencies.

Graph-based representations—such as control flow graphs (CFGs), program dependency graphs (PDGs), and data flow graphs (DFGs)—have gained prominence in code analysis tasks due to their capacity to encode the rich semantic relationships inherent in source code~\cite{zhang2025machine}. However, despite their representational power, existing graph-based approaches encounter two key limitations that hinder their effectiveness in tasks such as code clone detection.

\begin{itemize}
    \item Firstly, researchers mainly adopt message-passing mechanisms to capture node dependencies, such as GCN~\cite{liu2023learning, yuan2020local}, GAT~\cite{yu2023graph, li2020semantic}, GMN~\cite{wang2020detecting}, and GIN~\cite{zhang2024fsd}. However, these models predominantly capture local node interactions and struggle to effectively represent higher-order graph-level semantics or facilitate cross-graph comparisons—capabilities that are critical for accurately assessing code similarity.
    
    \item Secondly, individual graph representations typically emphasize specific aspects of code structure, such as control flow in CFGs or data dependencies in DFGs, but inherently neglect other critical aspects. To address this limitation, recent research has proposed hybrid representations that integrate multiple representations, aiming to achieve a more holistic understanding of code semantics. One pioneering example is the Flow-Augmented Abstract Syntax Tree (FA-AST) introduced by~\cite{wang2020detecting}, which enhances ASTs by explicitly incorporating flow-related information to better capture semantic similarity. Inspired by FA-AST, subsequent studies have explored alternative handcrafted fusion techniques by combining ASTs with additional graph structures such as CFGs~\cite{zhao2022precise, yuan2022java, liu2023learning}, DFGs~\cite{yuan2022java, liu2023learning}, and PDGs~\cite{jiang2022hierarchical}. Nevertheless, the efficacy of these manually engineered fusion methods remains uncertain. Due to a lack of rigorous ablation studies and systematic evaluations, recent analyses~\cite{zhang2025ast} have identified that such approaches are significantly overloaded and slower while only achieving marginal performance gains. 
\end{itemize}

Consequently, current methods face critical challenges: (i) inadequate exploration of graph information from multiple perspectives, particularly regarding long-range dependencies; and (ii) insufficient exploitation of complementary semantics encoded across diverse graph modalities.

To tackle these challenges, we propose the Multi-Graph Attentional Graph Neural Network for clone detection (MAGNET), a novel framework designed to leverage AST, CFG, and DFG inputs, employing attention mechanisms and GNNs to comprehensively capture code representations. Specifically, MAGNET comprises three core components: (i) a residual GNN combined with node self-attention to model both local and global (long-range) dependencies within each code graph; (ii) a selective gated cross-attention layer to effectively capture node-level cross-graph dependencies; and (iii) a Set2Set-based fusion mechanism to aggregate node embeddings from the AST, CFG, and DFG into a unified graph-level representation.

The contributions of this paper are summarized as follows:
\begin{itemize}
\item We introduce MAGNET, a multi-graph attentional framework designed explicitly for code clone detection, leveraging the complementary strengths of AST, CFG, and DFG representations.
\item We design a selective gated cross-attention mechanism that enables fine-grained modeling of inter-graph dependencies, along with a Set2Set-based fusion module that dynamically aggregates node representations into a unified graph-level embedding, allowing the model to adaptively integrate complementary information from the AST, CFG, and DFG.
\item We perform extensive experimental evaluations, demonstrating that MAGNET achieves state-of-the-art performance on standard benchmark datasets, significantly outperforming existing approaches.
\end{itemize}

The rest of this paper is organized as follows: Section~\ref{sec: background} provides background on code clone detection and relevant concepts. Section~\ref{sec:related_work} reviews related work, highlighting existing approaches and methodologies. Section~\ref{sec:Methodology} details the methodology of our proposed MAGNET framework. Section~\ref{sec:Experiment} presents our experimental setup and analyses results. Section~\ref{sec:Threats to Validity} discusses potential threats to validity. Finally, Section~\ref{sec:conclusion} concludes the paper and suggests directions for future research.

\section{Background}
\label{sec: background}
This section presents the background necessary for understanding our problem and proposed approach in three parts: code clone detection, code representations, and attention mechanisms.

\subsection{Code Clone Detection}

The objective of code clone detection is to identify code fragments that exhibit syntactic and/or semantic (functional) similarities. A widely adopted classification scheme divides code clones into four types based on their degree of similarity~\cite{bellon2007comparison}.

\begin{itemize}
\item \textbf{Type-1:} Code fragments that are identical except for superficial differences, such as variations in formatting or comments.
\item \textbf{Type-2:} Code fragments that exhibit minor modifications—such as changes in variable names, data types, or literals—while preserving syntactic structure. All differences permitted in Type-1 also apply.
\item \textbf{Type-3:} Code fragments with substantial structural edits, including added, removed, or reordered statements, but that retain similar core functionality. This category subsumes the differences observed in Type-1 and Type-2.
\item \textbf{Type-4:} Code fragments that implement the same functionality through entirely different syntactic structures.
\end{itemize}

While this classification is widely accepted, it remains largely qualitative. In particular, the boundary between Type-3 and Type-4 clones is often ambiguous, making it challenging to achieve consistent annotation and evaluation across clone detection systems.

To address this ambiguity, Svajlenko et al.~\cite{svajlenko2016bigcloneeval} introduced the BigCloneBench benchmark, which defines finer-grained subcategories of Type-3 and Type-4 clones based on ranges of syntactic similarity. These subcategories include:

\begin{itemize}
\item \textbf{Very Strongly Type-III (VST3):} 90–100\%.
\item \textbf{Strongly Type-III (ST3):} 70–90\%.
\item \textbf{Moderately Type-III (MT3):} 50–70\%.
\item \textbf{Weakly Type-III / Type-IV (WT3/4):} 0–50\%.
\end{itemize}

This study focuses on the detection of \emph{weakly Type-III and Type-IV} clones, also referred to as functional clones, which represent the most challenging category due to their low syntactic similarity and high semantic variation.

\subsection{Code representations}
In this study, we explore three widely adopted structural representations of source code that have been extensively utilized in the context of code clone detection: abstract syntax trees (ASTs), control flow graphs (CFGs), and data flow graphs (DFGs). Each representation captures a distinct dimension of code semantics:

\begin{itemize}
\item \textbf{Abstract Syntax Tree (AST):} ASTs provide a syntactic, tree-structured representation of source code, abstracting away superficial variations such as formatting or variable naming. They emphasize the hierarchical structure of code and are particularly effective at capturing syntactic similarities beyond surface-level tokens.
\item \textbf{Control Flow Graph (CFG):} CFGs model the order in which instructions are executed, capturing the logical flow of control through the program. This representation is essential for analyzing conditional branches, loops, and overall execution logic, making it suitable for identifying control-level similarities between code fragments.
\item \textbf{Data Flow Graph (DFG):} DFGs capture how data moves through a program by representing the dependencies between variables and operations. They are valuable for detecting clones that share similar computational behavior or data usage patterns.
\end{itemize}

Technically, these representations are inherently complementary: each reflects a different aspect of the underlying code semantics. Therefore, designing an effective hybrid representation scheme that can integrate these diverse views is crucial for building models capable of robustly detecting semantic and functional code clones.

\subsection{Attention Mechanism}

The \textit{self-attention} mechanism, formalized in the Transformer architecture~\cite{vaswani2017attention}, computes contextualized embeddings by modeling pairwise interactions between elements in an input sequence. Specifically, for an input sequence, three matrices—query ($Q$), key ($K$), and value ($V$)—are derived via learned linear projections. The attention output is calculated as:

\begin{equation}
\text{Attention}(Q, K, V) = \text{softmax}\left(\frac{QK^\top}{\sqrt{d_k}}\right)V
\end{equation}

where $d_k$ denotes the dimensionality of the key vectors, and the scaling factor $\sqrt{d_k}$ is introduced to stabilize gradients during training.

To capture diverse types of relationships, \textit{multi-head attention} applies self-attention in parallel across multiple subspaces, each with its own learned projection matrices:

\begin{equation}
\text{head}_i = \text{Attention}(QW_i^Q, KW_i^K, VW_i^V)
\end{equation}
\begin{equation}
h = \text{concat}(\text{head}_1, \ldots, \text{head}_n)W^O
\end{equation}

This allows the model to attend to different parts of the input with each head. The output from all heads is concatenated and linearly projected to obtain the final representation.

This architecture has shown remarkable success in various domains due to its capacity to capture both local and global dependencies effectively. Its variants have been extended to graph learning~\cite{velivckovic2017graph, yun2019graph， tan2023exploring}, computer vision~\cite{dosovitskiy2020image}, and beyond.

\section{Related work}
\label{sec:related_work}
With the increasing popularity of GNN-based networks in code clone detection, various code representations have been adopted to fit GNN-based architectures. Based on our review of the literature, existing representations used in GNN-based methods primarily fall into two categories: singular representations and fused representations. These categories are summarized as follows:

\subsection{Singular graph code representation}

Singular graph-based methods typically rely on a single type of graph to model code semantics. These methods are primarily based on abstract syntax trees (ASTs), program dependency graphs (PDGs), or control flow graphs (CFGs).

AST-based methods include the work of \cite{lu2021code}, which employed AST edges in combination with GCN and GAT, achieving better performance than traditional tree-based methods such as Tree-CNN and Tree-LSTM.

For approaches leveraging PDG and CFG, Mehrotra et al.~\cite{mehrotra2021modeling} introduced HOLMES, an attention-based Siamese Graph Neural Network that captures structured syntactic and semantic information from PDGs, demonstrating strong generalizability in semantic clone detection. Building on this direction, Yu et al.~\cite{yu2023graph} proposed a Siamese Graph Matching Network that incorporates attention mechanisms over CFG and PDG representations, achieving improvements in both semantic representation quality and computational efficiency.

In addition, Zhang et al.~\cite{zhang2023efficient} presented a model that combines AST with GCN and a Transformer-based architecture. This hybrid design effectively captures long-range semantic dependencies while maintaining low computational complexity, achieving state-of-the-art performance on standard benchmarks.

These singular representation approaches, while effective, inherently exhibit limitations in capturing the entirety of semantic information within the source code.

\subsection{Fusion graph code representation}

To overcome the limitations of singular representations, researchers have proposed fusion-based methods integrating multiple graph representations. One of the pioneering works in this domain is FA-AST by \cite{wang2020detecting}, which employs a Graph Neural Network (GNN) on flow-augmented abstract syntax trees to effectively capture syntactic and semantic similarities in code fragments. Following this idea, many researchers have tried to enhance AST-based representations by integrating the AST, CFG, and DFG. For instance, Zhao et al.~\cite{zhao2022precise} integrated the AST and CFG with hierarchical dependencies, employing GAT to improve detection accuracy. Similarly, Fang et al.~\cite{fang2020functional} and Xu et al.~\cite{xu2021sccd} have adopted this fusion methodology in their studies. Some studies have explored more comprehensive combinations of AST, CFG, and DFG. For example, Yuan et al.~\cite{yuan2022java} introduced an intermediate code-based graph representation that integrates these three components, thereby enhancing the identification of functional code clones. Additionally, Liu et al.~\cite{liu2023learning} proposed TAILOR, which incorporates AST, CFG, and DFG to improve the detection of functionally similar code fragments. Jiang et al.~\cite{jiang2022hierarchical} developed a Graph-LSTM framework that integrates hierarchical semantic features from both CFG and PDG into the AST, effectively capturing local syntactic and global semantic dependencies through sequential and topological ordering. Similar hybrid graph representations have also been extended to cross-language code clone detection, as demonstrated by \cite{mehrotra2023improving} and \cite{swilam2023cross}.

However, recent empirical analyses have raised critical concerns about the actual effectiveness of such hybrid graph representations. Zhang et al.~\cite{zhang2025ast} show that the performance gains from these manually engineered fusion methods are often marginal or inconsistent.

Another line of research attempts to perform the fusion of multiple code representations within the network architecture itself. To the best of our knowledge, this area remains largely unexplored. One of the few related efforts is the work by \cite{hua2020fcca}, who propose FCCA, a framework that concatenates features from text, AST, and CFG using separate neural network branches. However, their approach lacks architectural consistency, as each modality is processed using a distinct network design, making the fusion less principled and potentially unbalanced. Furthermore, their framework focuses solely on generating graph-level embeddings, overlooking fine-grained node-level interactions that are essential for capturing structural semantics.

Another partially related work is that of \cite{zhang2023efficient}, which introduces a cross-code attention mechanism applied over AST representations to capture semantic similarities between code fragments. While effective, their method is limited to a single graph modality and does not explore the fusion of multiple structural representations such as CFG or DFG.

\begin{figure*}[t]
    \centerline{\includegraphics[width = 1.\linewidth]{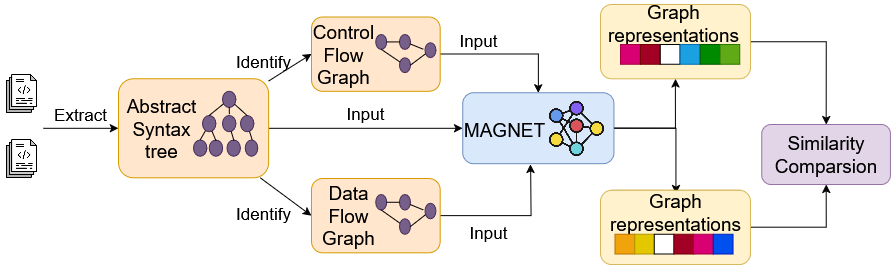}}
    \caption{Methodology Architecture Employed in Our Study.}
    \label{fig:methodology}
\end{figure*}

\section{Methodology}
\label{sec:Methodology}
In this paper, we propose a new framework called \textbf{MAGNET}, designed for multi-graph code understanding and pairwise similarity estimation in clone detection. An overview of the framework is presented in Figure~\ref{fig:methodology}. Given a pair of code functions, we first parse each function into an abstract syntax tree (AST). Based on the syntactic structure of the AST, we further derive the corresponding control flow graph (CFG) and data flow graph (DFG), capturing control logic and data dependencies, respectively. The resulting graphs are then fed into \textbf{MAGNET}, an attention-based graph similarity computation network. The detailed architecture of MAGNET is described in the following section. After processing by MAGNET, the fused graph embeddings of each code function are compared, and their cosine similarity is used to produce the final clone score.

To clearly present our approach, we organize this section into three parts. First, we formally define the code clone detection task as a graph-based similarity learning problem. Next, we describe the construction of multi-graph representations. Finally, we introduce the overall architecture of MAGNET.

\subsection{Problem Definition}
The code clone detection task in this study is formulated as a graph-based similarity learning problem. Given two code fragments, $C_i$ and $C_j$, each fragment is parsed into three structural graph representations: AST, CFG, and DFG. Accordingly, each code fragment $C_i$ is transformed into a set of graphs $\mathcal{G}_i = {G_i^{\text{AST}}, G_i^{\text{CFG}}, G_i^{\text{DFG}}}$, where each graph is defined as $G = (V, E, A)$, with $V$ denoting the set of nodes, $E$ the set of edges, and $A \in {0,1}^{|V| \times |V|}$ representing the adjacency matrix.

The objective is to compute a similarity score $s_{ij} \in [0, 1]$ between two sets of graphs, $\mathcal{G}_i$ and $\mathcal{G}_j$, that reflects the likelihood that the corresponding code fragments, $C_i$ and $C_j$, are semantically equivalent. A higher score implies stronger similarity, where:
\begin{itemize}
    \item $s_{ij} = 1$ indicates high semantic equivalence (e.g., functional clones);
    \item $s_{ij} = 0$ indicates no similarity.
\end{itemize}

To enable binary classification evaluation, we apply a threshold $\sigma \in (0, 1)$ to the predicted similarity score:
\begin{equation}
\hat{y}_{ij} =
\begin{cases}
1, & \text{if } s_{ij} > \sigma, \quad \text{(clone pair)} \\
0, & \text{otherwise,} \quad \text{(non-clone pair)}
\end{cases}
\end{equation}

\subsection{Code Graphs Construction}
To comprehensively capture the structural and semantic characteristics of source code, we employ a multi-graph representation approach. While traditional methods often rely on a singular representation (e.g., AST or CFG), such approaches are inherently limited in scope, capturing only specific aspects of program structure or behavior. In contrast, we utilize three complementary graph representations: the abstract syntax tree (AST), control flow graph (CFG), and data flow graph (DFG). Together, these representations offer a holistic view of each code fragment, encompassing both syntactic hierarchy and semantic execution and data dependencies.

These representations are language-agnostic and can be extracted from source code in various programming languages. In the following subsections, we describe the procedures for extracting each type of graph.

\subsubsection{Abstract Syntax Tree (AST)}
The abstract syntax tree (AST) provides a hierarchical representation of the syntactic structure of source code~\cite{baxter1998clone}. Formally, we define an AST as $G_{\text{AST}} = (V_{\text{AST}}, E_{\text{AST}})$, where $V_{\text{AST}}$ denotes the set of nodes representing syntactic constructs, and $E_{\text{AST}}$ denotes the set of directed edges capturing parent–child relationships among nodes.

In our implementation, we use the \texttt{javalang}~\cite{javalang} library to parse Java source code and extract ASTs. Each node in the resulting AST corresponds to a syntactic unit such as a method declaration, assignment, loop, or conditional expression. The \texttt{javalang} library abstracts each node to its structural role within the syntax tree, such as \texttt{MethodDeclaration}, \texttt{BinaryOperation}, or \texttt{IfStatement}. AST edges describe the nested structure of code—how high-level constructs are composed of or contain other constructs—thus capturing the grammatical hierarchy of the program.

While ASTs are effective in modeling syntactic composition, they lack the ability to capture execution semantics such as control or data dependencies. Consequently, ASTs alone are insufficient for detecting functional similarities and are therefore complemented by CFG and DFG representations in our framework.

\subsubsection{Control Flow Graph (CFG)}
The control flow graph (CFG) models the execution order of program statements by representing all possible paths that can be traversed during program execution~\cite{koppel2022automatically}. Following the approach of \cite{liu2023learning}, we construct control flow dependencies by deriving CFGs from the parsed abstract syntax tree (AST) through static analysis. This process identifies control constructs such as \texttt{if}, \texttt{while}, \texttt{switch}, and method invocations. The code is then segmented into basic blocks, each comprising a linear sequence of instructions without internal branching. New blocks are created at control constructs or jump targets, and directed edges are added to represent execution paths, including sequential edges between consecutive blocks, branching edges for conditionals, loop-back edges for iterations, and exit edges for returns or program termination.

\subsubsection{Data Flow Graph (DFG)}
Similar to CFG, nodes in a DFG correspond to program statements, but edges capture data dependencies, indicating how a variable’s value in one statement influences its use in another. To extract a DFG, we first identify variables defined and used in each AST statement, then apply reaching definitions analysis to connect definitions with their subsequent uses. These connections form use–def chains, from which we derive data flow edges.

\begin{figure*}[t]
    \centerline{\includegraphics[width = 1.\linewidth]{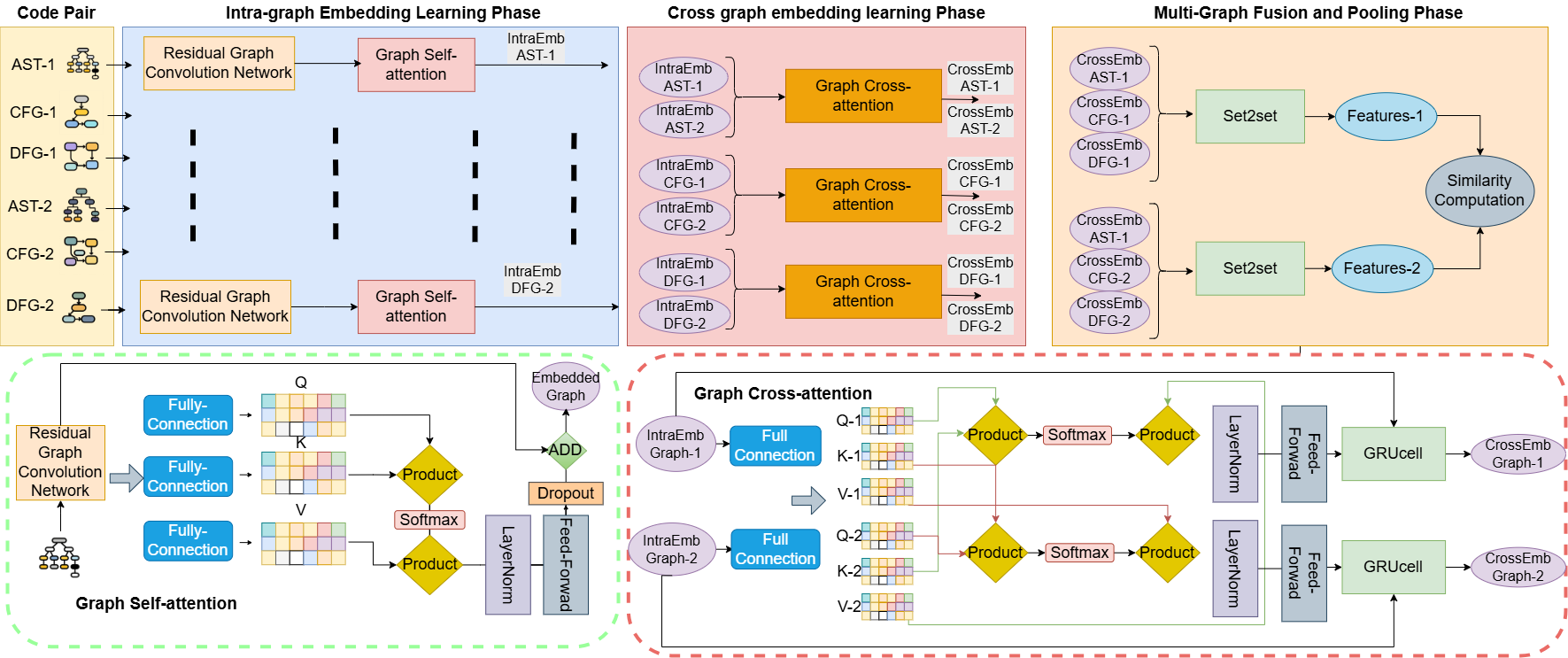}}
    \caption{Overview of MAGNET Network Structure.}
    \label{fig:network}
\end{figure*}

\subsection{Network Structure}

In this section, we present the architecture of our proposed model, \textbf{MAGNET}, illustrated in Figure~\ref{fig:network}. The model is designed to learn fine-grained semantic representations of code fragments by jointly leveraging multiple graph modalities. MAGNET consists of three main stages: (i) \textit{intra-graph embedding learning}, where local and global dependencies are captured within each individual graph (AST, CFG, and DFG); (ii) \textit{cross-graph embedding learning}, which models interactive relationships between paired code fragments through a gated cross-attention mechanism; and (iii) \textit{multi-graph fusion and pooling}, which integrates embeddings from different graph modalities into holistic, program-level representations for similarity computation. The final similarity score between two code fragments is then derived from these fused embeddings.

\subsubsection{Intra-graph Embedding Learning Phase}

In the first stage, the nodes of each input graph (AST, CFG, and DFG) are encoded into high-dimensional feature vectors to serve as inputs for downstream tasks. The goal is to capture both short-range structural dependencies and long-range semantic relationships within code graphs. To achieve this, we combine a residual graph convolutional network (GCN) layer, which aggregates local neighborhood information, with a node-level self-attention mechanism, which models global contextual interactions across all nodes.

\begin{itemize}
\item \textbf{Residual Graph Convolutional Network.}
GCNs are effective for learning local structural patterns, but deep GCNs often suffer from over-smoothing, where node representations become indistinguishable~\cite{li2018deeper}. To address this, we use a residual graph convolutional network that incorporates residual connections~\cite{bresson2017residual} between GCN layers and aggregates their outputs. Formally, for each layer $l \in \{1, \dots, L\}$:
\begin{equation}
H^{(l)} = \sigma \!\left( \hat{A} H^{(l-1)} W^{(l)} \right),
\end{equation}
where $\hat{A}$ is the normalized adjacency matrix, $W^{(l)} \in \mathbb{R}^{d \times d}$ is the learnable weight matrix at layer $l$, and $\sigma(\cdot)$ is the ReLU activation function. The residual aggregation is defined as:
\begin{equation}
H_{\text{GCN}} = \sum_{l=1}^{L} H^{(l)}.
\end{equation}

    \item \textbf{Node-level Self-attention.}  
    While residual graph convolution network block captures local neighborhood information, they struggle to represent long-range dependencies that are common in program graphs (e.g., variable declarations and uses in distant AST branches). To address this, we incorporate a node-level self-attention mechanism, enabling each node to directly attend to semantically relevant nodes across the entire graph.  

    Given the GCN output $H_{\text{GCN}} \in \mathbb{R}^{n \times d}$, we first normalize and compute query, key, and value projections:
    \begin{equation}
    \tilde{H} = \texttt{LN}(H_{\text{GCN}}), \quad 
    Q = \tilde{H}W_Q, \; K = \tilde{H}W_K, \; V = \tilde{H}W_V,
    \end{equation}
    where $W_Q, W_K, W_V \in \mathbb{R}^{d \times d}$ are learnable parameters.  

    The attention output is computed as:
    \begin{equation}
    A = \texttt{softmax}\!\left(\frac{QK^\top}{\sqrt{d}}\right), \quad
    O = AVW_O,
    \end{equation}
    with $W_O \in \mathbb{R}^{d \times d}$ as the output projection. A residual connection is then applied:
    \begin{equation}
    H_1 = \texttt{Dropout}(O) + H_{\text{GCN}}.
    \end{equation}

    Finally, to effectively integrate short-range structural information with long-range semantic dependencies, we apply a feed-forward network (\texttt{FFN}) with an additional residual connection:
    \begin{align}
    F &= \texttt{FFN}\!\left(\texttt{LN}(H_1)\right), \\
    H_{\text{ATT}} &= \texttt{Dropout}(F) + H_1.
    \end{align}
\end{itemize}

Overall, this stage integrates a residual graph convolutional network block, which preserves local structural features, with a self-attention stage, which captures global semantic relationships. The resulting node embeddings, $H_{\text{ATT}}$, provide a balanced representation that combines short- and long-range dependencies within code graphs, serving as a strong foundation for the cross-graph learning stage that follows.

\subsubsection{Cross graph embedding learning Phase}

Although the intra-graph stage captures local and global relationships within each AST, CFG, and DFG, accurate clone detection hinges on extracting cross-dependencies between the \emph{two} code fragments across these modalities~\cite{zhang2023efficient, wang2020detecting}. We therefore compute fine-grained, bidirectional correspondences at the node level with cross-code attention, and then apply a gated update to inject these pairwise signals while preserving each graph’s context.

To capture the interactions between code graphs, we utilize a cross-code attention mechanism. Formally, given node embeddings $H_{\text{ATT}1},\ H{\text{ATT}_2} \in \mathbb{R}^{n \times d}$ obtained from the intra-graph embedding learning stage (each derived from the AST, CFG, and DFG), we first normalize them:
\begin{equation}
\tilde{X}_1 = \texttt{LN}(H_{\text{ATT}_1}), \quad \tilde{X}_2 = \texttt{LN}(H_{\text{ATT}_2}).
\end{equation}

For $H_{\text{ATT}_1}$ attending to $H_{\text{ATT}_2}$:
\begin{equation}
Q_1 = \tilde{X}_1 W_Q, \; K_2 = \tilde{X}_2 W_K, \; V_2 = \tilde{X}_2 W_V,
\end{equation}
\begin{equation}
O_1 = \texttt{softmax}\!\left(\tfrac{Q_1 K_2^\top}{\sqrt{d}}\right) V_2 W_O.
\end{equation}

Similarly, for $H_{\text{ATT}_2}$ attending to $H_{\text{ATT}_1}$ we obtain $O_2$. The outputs are refined by a feed-forward network:
\begin{equation}
M_1 = \texttt{FFN}(\texttt{LN}(O_1)), \quad M_2 = \texttt{FFN}(\texttt{LN}(O_2)).
\end{equation}

Through the cross-graph co-attention mechanism, directional similarity information between the two code graphs is explicitly retained, enriching the representations with fine-grained interaction signals. Then, we update the final node states of the cross-graphs with a recurrent gating function (\texttt{GRUCell}):

\begin{equation}
H_1 = \texttt{GRUCell}(M_1, H_{\text{ATT}_1}), \quad 
H_2 = \texttt{GRUCell}(M_2, H_{\text{ATT}_2}).
\end{equation}
This gated update is capable of preserving the original embedding context while selectively incorporating cross-graph information. It is worth noting that our framework adopts a Siamese architecture; therefore, all parameters in both the intra-graph embedding and cross-graph embedding stages are shared across all input graphs, including the paired code fragments as well as the different graph modalities (AST, CFG, and DFG).

\subsubsection{Multi-Graph Fusion and Pooling Phase}
To obtain program-level representations from multiple variable-sized graph embeddings, we employ a Set2Set pooling layer~\cite{vinyals2015order} that aggregates node embeddings into a unified attention space. The iterative LSTM controller in Set2Set allows the pooling process to capture cross-modal interactions beyond simple averaging. Furthermore, because Set2Set is permutation-invariant and independent of graph size or node ordering, it can effectively merge variable-sized graphs into a coherent program representation while preserving the unique contribution of each modality.

Formally, given node embeddings ${h_i}_{i=1}^N$, Set2Set maintains a query vector $q_t$, updated by an LSTM. At each iteration $t$:
\begin{align}
q_t &= \texttt{LSTM}(q^{*}_{t-1}), \label{eq:set2set_q}\\
\alpha_{i,t} &= \texttt{softmax}_i \!\left( h_i \cdot q_t \right), \label{eq:set2set_alpha}\\
r_t &= \sum_{i=1}^N \alpha_{i,t}\, h_i, \label{eq:set2set_r}\\
q^{*}_t &= [\, q_t \,\|\, r_t \,], \label{eq:set2set_concat}
\end{align}
where $q_t$ is the query state produced by the LSTM, $\alpha_{i,t}$ represent the attention weights over nodes, $r_t$ is the attention readout, and $q^{*}_t$ concatenates the query and the readout.

In our framework, we concatenate (\texttt{concat}) the embeddings from the AST, CFG, and DFG for each program:
\begin{align}
H^{(1)} &= \texttt{concat}\!\left(H^{\text{AST}}_1,\, H^{\text{CFG}}_1,\, H^{\text{DFG}}_1\right), \\
H^{(2)} &= \texttt{concat}\!\left(H^{\text{AST}}_2,\, H^{\text{CFG}}_2,\, H^{\text{DFG}}_2\right),
\end{align}
and apply \texttt{Set2Set} pooling to obtain the global program representations:
\begin{equation}
h^{(1)} =\texttt{Set2Set}\!\left(H^{(1)}\right), \quad
h^{(2)} = \texttt{Set2Set}\!\left(H^{(2)}\right).
\end{equation}

The resulting vectors, $h^{(1)}$ and $h^{(2)}$, serve as fused program embeddings that integrate syntactic, control-flow, and data-flow information and are used for downstream similarity computation.

\subsubsection{Calculating code similarity}
Using the pooled vectors, $h^{(1)}$ and $h^{(2)}$, we calculate the similarity using the cosine similarity function:
\begin{equation}
s = \cos\!\left(h^{(1)}, h^{(2)}\right) = 
\frac{h^{(1)} \cdot h^{(2)}}{\|h^{(1)}\| \, \|h^{(2)}\|},
\end{equation}
where $s \in [-1, 1]$ denotes the similarity score between the two program representations.

For loss calculation, we compare the predicted similarity $s$ with the ground-truth label $\tilde{y} \in {-1, 1}$ using the mean squared error (MSE) loss:
\begin{equation}
\mathcal{L} = \texttt{MSELoss}(s, \tilde{y}) 
= \frac{1}{N} \sum_{i=1}^N \left(s_i - \tilde{y}_i\right)^2,
\end{equation}

\section{Experimental settings}

\subsection{Dataset}
We evaluate our model on BigCloneBench~\cite{svajlenko2014towards} and GoogleCodeJam~\cite{googlecodejam}. 

\begin{itemize}
\item \textbf{BigCloneBench.} This dataset is a widely used, large-scale code clone benchmark containing over 6,000,000 true clone pairs and 260,000 false clone pairs across 10 different functionalities. All code in BigCloneBench is written in Java. The dataset categorizes semantic code clones based on statement-level similarity scores in the range [0, 1): strongly Type-3 (ST3) with similarity [0.7, 1.0), moderately Type-3 (MT3) with similarity [0.5, 0.7), and weakly Type-3/Type-4 (WT3/T4) with similarity [0.0, 0.5).

Table~\ref{tab:clone_distribution} summarizes the distribution of clone types in BigCloneBench. As the majority of clone pairs are weak Type-3/Type-4, the dataset is well-suited for evaluating semantic clone detection. We use the dataset setup provided by FA-AST~\cite{wang2020detecting}, which removes fragments without any labeled true or false clone pairs and balances the number of true and false clones in the training set.
\begin{table}[h]
\centering
\caption{Distribution of clone types in the BigCloneBench dataset.}
\label{tab:clone_distribution}
\begin{tabular}{lcccccc}
\hline
\textbf{Clone Type} & T1 & T2 & VST3 & ST3 & MT3 & T4 \\
\hline
\textbf{Percentage (\%)} & 0.28 & 0.06 & 0.04 & 0.18 & 0.96 & 98.47 \\
\hline
\end{tabular}

\end{table}

\item \textbf{Google Code Jam.} This dataset consists of 1,669 Java files drawn from 12 different competition problems. Prior work has confirmed that only a small fraction of clone pairs in this dataset are syntactically similar~\cite{zhao2018deepsim}. Therefore, we assume that the clone pairs in Google Code Jam are primarily Type-4 clones. Following the dataset setting in~\cite{wang2020detecting}, we construct the Google Code Jam dataset accordingly. Following prior work~\cite{jia2025code, zhao2018deepsim}, we randomly select 1,000 code snippets for training, 332 for validation, and 32 for testing, yielding 499,500 training pairs, 54,946 validation pairs, and 54,946 testing pairs.
\end{itemize}

Table~\ref{tab:demographics_dataset} shows the basic information for the two datasets used in our experiments.

\begin{table}[]
\centering
\caption{demographics of GoogleCodeJam and BigCloneBench}
\label{tab:demographics_dataset}
\begin{tabular}{lll}
\hline
                       & GoogleCodeJam & BigCloneBench \\ \hline
Code Fragments         & 1664   &  9133  \\
Average lines of code  & 52.97   &  32.79  \\
Vocabulary size        & 8,033   & 77,535     \\ 
True clone pairs       & 121,130   & 336,498    \\
False Clone pairs      & 488,262   &  947,020  \\ \hline
\end{tabular}
\end{table}

\subsection{Implementation details}

We implement the models using PyTorch~\cite{paszke2019pytorch} and PyTorch Geometric~\cite{fey2019fast}. All experiments are conducted on a machine equipped with an Intel i9-13900K CPU, 32 GB of RAM, and an NVIDIA RTX A4000 with 16 GB of memory. We follow the dataset setting of FA-AST~\cite{wang2020detecting}, splitting the datasets into approximately 60\%, 20\%, and 20\% for training, validation, and testing, respectively. The model is trained for 5 epochs with a batch size of 10 and an initial learning rate of $5 \times 10^{-4}$ using the Adam optimizer. A learning rate reduction strategy is applied with a patience of 1 epoch and a decay factor of 0.5.

The default hyperparameter configuration of the proposed network is as follows:
\begin{itemize}
    \item \textbf{Intra-graph Embedding Learning:}  
The number of GNN layers is set to $L = 3$, with node embeddings of dimension $d = 128$. The self-attention module uses 8 heads, each with a dimension of 64, and applies a dropout rate of 0.1.

    \item \textbf{Cross-graph Embedding Learning:}  
The embedding dimension is fixed at 128, with 8 attention heads and a head dimension of 64.

    \item \textbf{Graph-level Pooling:}  
The Set2Set pooling layer uses an embedding dimension of 128 and runs for $T = 3$ iterative steps.
\end{itemize}

\subsection{Research Questions}
Given the proposed detection framework, this study aims to address the following research questions:

\begin{itemize}
\item \textbf{RQ1: How does the proposed MAGNET framework perform compared to state-of-the-art methods in code clone detection?}

This question aims to evaluate the overall effectiveness of MAGNET by benchmarking its performance against leading existing approaches on standard code clone detection datasets.

\item \textbf{RQ2: Does the integration of multi-graph fusion (AST, CFG, DFG) enhance the performance of code clone detection?}  

This question aims to explore whether the fusion of complementary structural representations enhances the model’s ability to capture semantic and functional similarities between code fragments.

\item \textbf{RQ3: To what extent do node-level self-attention and gated cross-attention contribute to model performance?}  
This question investigates the individual and combined impacts of the proposed attention mechanisms on capturing intra- and inter-graph semantic dependencies.

\item \textbf{RQ4: How efficient is MAGNET compared to other code clone detection tools?} 
This question examines the computational efficiency of MAGNET by assessing its scalability, training cost, and inference speed.
\end{itemize}

\section{Experimental analyses and results}
\label{sec:Experiment}
In this section, we present and analyze the results of each research question.

\subsection{Effectiveness of code clone detection (RQ1)}

To assess the effectiveness of our model, we conduct a comprehensive evaluation of MAGNET in comparison with other state-of-the-art clone detection tools, categorized into token-based, AST-based, and graph-based approaches.
\begin{enumerate}
    \item Token-based 
     \begin{itemize}
         \item \textbf{SourcererCC~\cite{sajnani2016sourcerercc}:} A tool that identifies code clones by computing the overlap of indexed token subsequences.
         \item \textbf{CodeBERT~\cite{feng2020codebert}:} A pre-trained Transformer model on source code and natural language that encodes token sequences into semantic embeddings for clone detection tasks.
         \item \textbf{AlphaCC~\cite{jia2025code}:} An AlphaFold-inspired framework that represents code fragments as token sequences and constructs a multiple sequence alignment (CodeMSA) from lexically similar sequences. It employs a modified attention-based encoder to capture both intra-sequence and cross-sequence dependencies.
     \end{itemize}
     
    \item AST-based
    \begin{itemize}
        \item \textbf{TBCCD~\cite{yu2019neural}:} An approach that utilizes tree-based convolution over token-enhanced ASTs to detect semantic code clones.
        \item \textbf{ASTNN~\cite{zhang2019novel}:} A state-of-the-art model that splits large ASTs into sequences of small statement trees and encodes them into lexical and syntactic vectors for similarity comparison. 
        \item \textbf{Code-Token-Learner~\cite{zhang2023efficient}:} A Transformer-based model designed to handle long code fragments by first compressing AST representations through a GCN-based token learner, which aggregates nodes into a compact set of informative tokens.
    \end{itemize}
    
    \item Graph-based
    \begin{itemize}
        \item \textbf{FA-AST + GMN~\cite{wang2020detecting}:} A state-of-the-art model that augments ASTs with explicit control- and data-flow edges, followed by Graph Matching Networks (GMN) to compute the similarity of code pairs.
        \item \textbf{PNIAT+CFG and PNIAT+PDG~\cite{yu2023graph}:} A technique designed to convert irregular code graph data and use attention mechanisms to capture critical tokens within each statement in parallel for semantic clone detection.
    \end{itemize}
\end{enumerate}

The results on BigCloneBench across different clone types are presented in Table~\ref{tab:performance_comparison}.
As shown, all clone detection methods perform well on Type-1 and Type-2 clones, except for the non-ML baseline (SourcererCC), which lags significantly behind its ML-based counterparts. However, for semantic clones (MT3 and T4), even advanced ML-based detectors experience a notable performance drop compared to their near-perfect scores on T1 and T2, indicating that generalizing to semantic equivalence remains challenging.

In contrast, MAGNET achieves near-perfect recall across all clone categories, reaching 100\% on Type-1 to ST3 and maintaining 99.3\% and 96.7\% recall on MT3 and T4 clones, respectively. This demonstrates the model’s ability to capture functional similarity beyond surface-level patterns. While AlphaCC attains the highest precision (97.7\%), MAGNET strikes a stronger balance between precision (96.9\% — accuracy of positive code clone predictions) and recall (96.1\% — ability to find all actual code clones), resulting in the best overall F1 score (96.5%).

We further validate the effectiveness of MAGNET on the Google Code Jam dataset, which is predominantly composed of semantic (Type-4) clones. As shown in Table~\ref{tab:performance_gcj}, our model achieves an F1 score of 99.2\%, clearly surpassing all state-of-the-art methods. In comparison, the strongest competitors—AlphaCC (97.4\%), Code-Token-Learner (96.0\%), and ASTNN (95.4\%)—lag behind by a notable margin, confirming that MAGNET generalizes more effectively to challenging semantic clone detection tasks.

\begin{table*}[!htpb]
\centering
\caption{Code clone detection results on BigCloneBench}
\label{tab:performance_comparison}
\begin{tabular}{p{1.5cm}p{3.0cm}p{0.9cm}p{0.9cm}p{0.9cm}p{0.9cm}p{0.9cm}p{0.9cm}p{1.4cm}p{0.8cm}}
\hline
\multirow{2}{*}{Type}        & \multirow{2}{*}{Method} & \multicolumn{6}{c}{Recall}                & \multirow{2}{*}{Precision} & \multirow{2}{*}{F1} \\ \cline{3-8}
                             &                         & T1  & T2   & ST3  & MT3  & T4   & Overall &                          &          \\ \hline
\multirow{3}{*}{Token-based} & SourcererCC             & 100 & 98.0 & 61.0 & 5.0  & 0    & 1.0     & 88.0      & 1.0      \\
                             & CodeBERT                & 100 & 100  & 99.8 & 97.0 & 93.2 & 93.4    & 94.7      & 94.1     \\
                             & AlphaCC                 & 100 & 100  & 100  & 100  & 93.8 & 94.2    & \textbf{97.7}      & 96.0     \\ \hline
\multirow{3}{*}{AST-based}   & TBCCD                   & 100 & 100  & 100  & 98.1 & 93.0 & 93.1    & 96.6      & 94.8     \\
                             & ASTNN                   & 100 & 100  & 99.6 & 97.9 & 92.2 & 92.3    & 93.4      & 92.9     \\
                             & Code-Token-Learner      & 100 & 100  & 100  & 98.5 & 94.8 & 95.2    & 96.1      & 95.6     \\ \hline
\multirow{4}{*}{Graph-based} & FA-AST + GMN            & 100 & 100  & 96.6 & 96.5 & 93.5 & 93.6    & 95.8      & 94.7     \\
                             & PNIAT+CFG               & 100 & 100  & 100  & 99.2 & 96.4 & 96.6    & 92.4      & 94.5     \\
                             & PNIAT+PDG               & 100 & 100  & 99.2 & 98.9 & 91.8 & 92.0    & 95.9      & 93.8     \\
                             & \textbf{MAGNET (Ours)} & 100 & 100 & 100 & \textbf{100} & \textbf{97.0} & \textbf{96.9} & 96.1 & \textbf{96.5} \\ \hline
\end{tabular}
\end{table*}

\begin{table*}[!htbp]
\centering
\caption{Code clone detection results on GoogleCodeJam}
\label{tab:performance_gcj}
\begin{tabular}{lllll}
\hline
Type                         & Method             & Recall & Precision & F1 \\ \hline
\multirow{3}{*}{Token-based} & SourcererCC        & 4.2    & 74.8      & 7.9      \\
                             & CodeBERT           & 94.0   & 94.5      & 94.2     \\
                             & AlphaCC            & 97.2   & 97.6      & 97.4     \\ \hline
\multirow{3}{*}{AST-based}   & TBCCD              & 92.7   & 97.3      & 95.0     \\
                             & ASTNN              & 93.0   & 97.9      & 95.4     \\
                             & Code-Token-Learner & 95.8   & 96.3      & 96.0     \\ \hline
\multirow{4}{*}{Graph-based} & FA-AST + GMN       & 95.0   & 94.6      & 94.8     \\
                             & PNIAT+CFG          & 94.9   & 93.4      & 94.1     \\
                             & PNIAT+PDG          & 96.0   & 93.8      & 94.9     \\
                             & \textbf{MAGNET (Ours)} & \textbf{99.6} & \textbf{98.7} & \textbf{99.2} \\ \hline
\end{tabular}
\end{table*}

\subsection{Evaluation of multi-graph fusion (RQ2)}

To evaluate the effectiveness of our multi-graph fusion strategy and the contribution of each graph representation to clone detection, we conduct experiments using single-graph inputs (AST, CFG, DFG), as well as their pairwise and full combinations, as inputs to our network on BigCloneBench. The results are presented in Table~\ref{tab:rq2}.

We can see from Table~\ref{tab:rq2} that all single-graph and combined-graph settings achieve perfect recall for Type-1 and Type-2 clones. However, for semantic clones, any configuration that excludes the AST (i.e., CFG, DFG, or CFG+DFG) performs poorly. This suggests that flow-based semantics alone are insufficient for clone detection and that the syntactic information provided by the AST is always required as a foundation.

AST alone achieves relatively strong performance (F1 = 95.6\%), reflecting its effectiveness in capturing the syntactic backbone of programs. When combined with the AST, both the CFG and DFG provide complementary improvements: AST+CFG reaches an F1 of 95.8\%, while AST+DFG achieves an F1 of 96.1\%. These results demonstrate that, within our proposed framework, control-flow and data-flow information effectively enrich the syntactic structure and enhance detection performance. With all three representations fused, MAGNET achieves the best overall performance (F1 = 96.5\%), surpassing all single- and dual-graph variants. This confirms the effectiveness of our fusion strategy and shows that integrating the AST, CFG, and DFG enables the model to more comprehensively capture semantic equivalence.

\begin{table*}[!htbp]
\centering
\caption{Evaluation results of our multi-graph fusion stretegy on BigCloneBench.}
\label{tab:rq2}
\begin{tabular}{lllllllll}
\hline
\multirow{2}{*}{Graph Combinations} & \multicolumn{6}{l}{Recall}         & \multirow{2}{*}{Precision} & \multirow{2}{*}{F1} \\ \cline{2-7}
                                    & T1 & T2 & ST3 & MT3 & T4 & Overall &                            &                           \\ \hline
AST                                 & 100  & 100  & 100   & 99.1   & 95.7  & 95.8       & 95.4                          & 95.6                         \\
CFG                                 & 100  & 100  & 100   & 72.9   & 46.3  & 47.0       & 34.1                          & 39.5                         \\
DFG                                 & 100  & 100  &  100  & 88.1   & 64.1  & 53.2        & 25.2                         & 34.2                         \\
AST + CFG                           & 100  & 100  & 100   & 100   & 96.6  & 96.7       & 95.0                          & 95.8                         \\
AST + DFG                           & 100  & 100  & 100   & 100   & 95.7  & 95.9      & \textbf{96.3}                          & 96.1                         \\
CFG + DFG                           & 100  & 100  & 100   & 72.7   & 49.8  & 50.4       & 34.0                         & 40.6                         \\ \hline
AST + CFG +DFG (MAGNET)             & 100  & 100  & 100   & 100   & \textbf{97.0}  & \textbf{96.9}       & 96.1                          &  \textbf{96.5}                        \\ \hline
\end{tabular}
\end{table*}

\subsection{Ablation study (RQ3)}

To validate the effectiveness of the key components of our network, we conduct an ablation study on BigCloneBench. The results are shown in Table~\ref{tab:rq3}. Specifically, we evaluate four ablation settings in comparison with the full model:

\begin{itemize}
\item \textbf{No Residuals.} We replace the residual GNN blocks with vanilla GNN layers so that intra-graph embeddings are produced solely by stacked GNNs without residual aggregation.
\item \textbf{No Intra-Attention.} We disable the self-attention phase in intra-graph embedding learning, relying only on residual GNNs. This isolates the contribution of long-range dependency modeling across nodes within each AST, CFG, and DFG.
\item \textbf{No Cross-Attention.} We remove the cross-graph co-attention and GRU-based gated updates. Instead, the two code fragment embeddings are concatenated and passed through a fully connected layer to obtain the prediction, without any explicit node-level interaction. 
\item \textbf{No Set2Set.} We replace the Set2Set pooling layer with alternative pooling strategies (i.e., mean pooling and global attention pooling). Among these, the best-performing variant (global attention pooling) is reported as the ablation result.
\end{itemize}

As shown in Table~\ref{tab:rq3}, removing any single component leads to a decline in overall performance compared with the full model. Among all variants, eliminating residual connections causes the largest degradation (F1 = 94.2, –2.38\%), underscoring their importance in mitigating over-smoothing and preserving discriminative node features, particularly for complex graph structures. Disabling intra-attention also reduces performance (F1 = 94.7, –1.87\%), highlighting the critical role of long-range dependency modeling in capturing semantic relations across distant nodes within the AST, CFG, and DFG.  

Interestingly, removing cross-attention results in a smaller drop (F1 = 95.2, –1.35\%) compared with intra-attention, suggesting that modeling long-range dependencies within graphs is more influential than capturing inter-graph dependencies. Finally, replacing Set2Set pooling with alternative strategies leads to a decline (F1 = 94.8, –1.76\%), confirming that attention-based iterative pooling is more effective at integrating multi-graph embeddings than simple averaging or global pooling.

\begin{table*}[!htbp]
\centering
\caption{Ablation study on BigCloneBench showing the performance impact of removing key components from Model on BigCloneBench.}
\label{tab:rq3}
\begin{tabular}{lp{1cm}p{1cm}lllllllll}
\hline
\multicolumn{4}{l}{\multirow{1}{*}{Components}}        & \multicolumn{6}{l}{Recall} & \multirow{2}{*}{Precision} & \multirow{2}{*}{F1} \\ \cline{1-10}
Residual & Intra-attention & Cross-attention & Set2Set & \multirow{1}{*}{T1} & \multirow{1}{*}{T2} & \multirow{1}{*}{ST3} & \multirow{1}{*}{MT3} & \multirow{1}{*}{T4} & \multirow{1}{*}{Overall} &                            &                           \\ \hline
\ding{55}        & \checkmark               & \checkmark               & \checkmark       & 100                   & 100                   & 100                    & 100                    & 98.5                   & 98.5                        & 90.4                          & 94.2                         \\
\checkmark        & \ding{55}              & \checkmark               & \checkmark       & 100                  & 100                  & 100                    & 99.8                    & 96.6                   & 96.7                        & 92.8                          & 94.7                         \\
\checkmark        & \checkmark               & \ding{55}              & \checkmark       & 100                   & 100                   & 100                    & 99.3                    & 94.4                    & 94.6                        & 95.7                          & 95.2                         \\
\checkmark        & \checkmark               & \checkmark               & \ding{55}       & 100                  & 100                   & 99.3                    & 99.0                    & 95.2                   & 95.3                        & 94.3                          & 94.8                         \\
\checkmark         & \checkmark                & \checkmark                & \checkmark        & 100                   & 100                   & 100                    & 100                    & 97.0                   & 96.9                        & 96.1                         & 96.5                         \\  \hline
\end{tabular}
\end{table*}

%%%%%%%%%%%%%%%%%%%%%%%%%
%%%%%%%%%%%%%%%%%%%%%%%%%
\subsection{Computational Efficiency (RQ4)}

\begin{table*}[!htbp]
\centering
\caption{Computational Efficiency of different methods on BigCloneBenc and GoogleCodeJam}
\label{tab:efficiency}
\begin{tabular}{lllp{1.5cm}p{2cm}lp{1.5cm}p{2cm}}
\hline
\multicolumn{2}{l}{Datasets}                  & \multicolumn{3}{l}{BigCloneBench}          & \multicolumn{3}{l}{GoogleCodeJam}      \\ \hline
\multicolumn{2}{l}{Metrics}                   & F1 & Training Time (s) & Prediction Time (s) & F1   & Training Time (s) & Prediction Time (s)\\ \hline
\multirow{5}{*}{Methods} & FA-AST + GMN       & 94.7     & 6701.12      & 2099.66        & 94.8 & 1872.98             & 363.3               \\
                         & SourcererCC        & 1.0      & -             & 19             & 7.9  & -             &   7             \\
                         & CoderBERT          & 94.1     & -             & 7368.34              & 94.2 & -             & 1837.6               \\
                         & Code-Token-Learner & 95.6     & 7751.52      & 1078.59       & 96.0 & 2437.56             &   116.8             \\
                         & MAGNET          & 96.5     & 7335.57      & 3192.29        & 99.2 &  2802.69           & 576.3                 \\ \hline
\end{tabular}
\end{table*}

To investigate this research question, we evaluate MAGNET’s computational efficiency compared with other state-of-the-art methods using BigCloneBench as the benchmark. Following prior work~\cite{dou2024cc2vec}, we divide the efficiency analysis into the training phase and the prediction phase, which reflect real-world usage scenarios. The results are shown in Table~\ref{tab:efficiency}.

Compared with existing approaches, MAGNET shows lower efficiency in both training and inference. This is due to the additional cost of processing multiple code modalities (AST, CFG, DFG), as well as the attentional mechanisms that operate at both intra- and inter-graph levels. While these components significantly enhance detection accuracy, they introduce computational overhead. As in real-world scenarios, there is a trade-off to be struck between efficiency and accuracy (we want quick and accurate models); this might render our approach impractical for some use cases—although it remains more computationally efficient than CodeBERT.
As future work, we plan to mitigate this overhead by incorporating lightweight attention mechanisms, graph sampling or pruning, and adaptive modality fusion to improve scalability and inference speed.

\section{Threats to Validity}
\label{sec:Threats to Validity}

We address several validity threats in this study. The first relates to efficiency. As outlined earlier, the primary objective of this work is to investigate the fusion of multiple code modalities and the proposed attention mechanisms. While our approach demonstrates the effectiveness of both the fusion strategy and the attention framework, these enhancements inevitably introduce additional computational overhead. Consequently, the current efficiency of our model falls short of most state-of-the-art methods, with the exception of CodeBERT.

A second concern involves generalizability. The reliance on multiple graph modalities poses challenges for broader application. Unlike token-based or AST-based approaches, graph-based methods require specialized tools to construct code graphs, and such tools are not yet standardized across programming languages. This limitation restricts the adaptability of our method to diverse languages and scenarios.

The limited generalizability of our approach also raises challenges for language diversity in evaluation. Our experiments were conducted on two widely used Java datasets—BigCloneBench and Google Code Jam. Although other clone detection datasets exist in different languages (e.g., OJClone~\cite{mou2016convolutional}), constructing multi-graph representations for such datasets is considerably more complex and time-intensive, which constrained our evaluation scope.

\section{Conclusion}
\label{sec:conclusion}
In this study, we presented MAGNET, a multi-graph attention framework that integrates AST, CFG, and DFG representations to enhance the performance of code clone detection. To achieve this, we adopt residual GNN layers for robust local structure learning, a node-level self-attention phase to model long-range dependencies, a gated cross-attention mechanism to capture fine-grained inter-graph interactions between code pairs, and Set2Set pooling to fuse modality-specific embeddings into a unified program representation.

Extensive experiments on BigCloneBench and Google Code Jam show that MAGNET delivers strong, often state-of-the-art, results. Ablation analyses confirm that each architectural component contributes materially to performance; additional ablations on the fusion strategy (single-, pairwise-, and tri-graph inputs) demonstrate the effectiveness of multi-graph integration and quantify the distinct contributions of each modality (i.e., AST, CFG, and DFG).

Currently, we validate the effectiveness of multi-graph fusion and attention mechanisms, which introduce additional computational overhead. Future work will prioritize efficiency—particularly lightweight or sparse attention, graph pruning or sampling, and adaptive modality selection.

\section*{Acknowledgment}
This publication has emanated from research conducted with the financial support of Taighde Éireann – Research Ireland under Grant Numbers 18/CRT/6223 and 13/RC/2094\_2.

\bibliographystyle{IEEEtran}
\bibliography{ref}

% Generated by IEEEtran.bst, version: 1.14 (2015/08/26)
\begin{thebibliography}{10}
\providecommand{\url}[1]{#1}
\csname url@samestyle\endcsname
\providecommand{\newblock}{\relax}
\providecommand{\bibinfo}[2]{#2}
\providecommand{\BIBentrySTDinterwordspacing}{\spaceskip=0pt\relax}
\providecommand{\BIBentryALTinterwordstretchfactor}{4}
\providecommand{\BIBentryALTinterwordspacing}{\spaceskip=\fontdimen2\font plus
\BIBentryALTinterwordstretchfactor\fontdimen3\font minus \fontdimen4\font\relax}
\providecommand{\BIBforeignlanguage}[2]{{%
\expandafter\ifx\csname l@#1\endcsname\relax
\typeout{** WARNING: IEEEtran.bst: No hyphenation pattern has been}%
\typeout{** loaded for the language `#1'. Using the pattern for}%
\typeout{** the default language instead.}%
\else
\language=\csname l@#1\endcsname
\fi
#2}}
\providecommand{\BIBdecl}{\relax}
\BIBdecl

\bibitem{saini2018code}
N.~Saini, S.~Singh \emph{et~al.}, ``Code clones: Detection and management,'' \emph{Procedia computer science}, vol. 132, pp. 718--727, 2018.

\bibitem{roy2007survey}
C.~K. Roy and J.~R. Cordy, ``A survey on software clone detection research,'' \emph{Queen’s School of computing TR}, vol. 541, no. 115, pp. 64--68, 2007.

\bibitem{tsantalis2017clone}
N.~Tsantalis, D.~Mazinanian, and S.~Rostami, ``Clone refactoring with lambda expressions,'' in \emph{2017 IEEE/ACM 39th International Conference on Software Engineering (ICSE)}.\hskip 1em plus 0.5em minus 0.4em\relax IEEE, 2017, pp. 60--70.

\bibitem{tsantalis2015assessing}
N.~Tsantalis, D.~Mazinanian, and G.~P. Krishnan, ``Assessing the refactorability of software clones,'' \emph{IEEE Transactions on Software Engineering}, vol.~41, no.~11, pp. 1055--1090, 2015.

\bibitem{nishi2018scalable}
M.~A. Nishi and K.~Damevski, ``Scalable code clone detection and search based on adaptive prefix filtering,'' \emph{Journal of Systems and Software}, vol. 137, pp. 130--142, 2018.

\bibitem{li2006cp}
Z.~Li, S.~Lu, S.~Myagmar, and Y.~Zhou, ``Cp-miner: Finding copy-paste and related bugs in large-scale software code,'' \emph{IEEE Transactions on software Engineering}, vol.~32, no.~3, pp. 176--192, 2006.

\bibitem{ebrahimi2019hmm}
N.~Ebrahimi, A.~Trabelsi, M.~S. Islam, A.~Hamou-Lhadj, and K.~Khanmohammadi, ``An hmm-based approach for automatic detection and classification of duplicate bug reports,'' \emph{Information and Software Technology}, vol. 113, pp. 98--109, 2019.

\bibitem{fokam2021influence}
M.~A. Fokam and R.~Ajoodha, ``Influence of contrastive learning on source code plagiarism detection through recursive neural networks,'' in \emph{2021 3rd International Multidisciplinary Information Technology and Engineering Conference (IMITEC)}.\hskip 1em plus 0.5em minus 0.4em\relax IEEE, 2021, pp. 1--6.

\bibitem{cheers2020novel}
H.~Cheers and Y.~Lin, ``A novel graph-based program representation for java code plagiarism detection,'' in \emph{Proceedings of the 3rd international conference on software engineering and information management}, 2020, pp. 115--122.

\bibitem{xiao2017evolution}
G.~Xiao, Z.~Zheng, and H.~Wang, ``Evolution of linux operating system network,'' \emph{Physica A: Statistical Mechanics and its Applications}, vol. 466, pp. 249--258, 2017.

\bibitem{sun2021vdsimilar}
H.~Sun, L.~Cui, L.~Li, Z.~Ding, Z.~Hao, J.~Cui, and P.~Liu, ``Vdsimilar: Vulnerability detection based on code similarity of vulnerabilities and patches,'' \emph{Computers \& Security}, vol. 110, p. 102417, 2021.

\bibitem{sajnani2016sourcerercc}
H.~Sajnani, V.~Saini, J.~Svajlenko, C.~K. Roy, and C.~V. Lopes, ``Sourcerercc: Scaling code clone detection to big-code,'' in \emph{Proceedings of the 38th international conference on software engineering}, 2016, pp. 1157--1168.

\bibitem{bellon2007comparison}
S.~Bellon, R.~Koschke, G.~Antoniol, J.~Krinke, and E.~Merlo, ``Comparison and evaluation of clone detection tools,'' \emph{IEEE Transactions on software engineering}, vol.~33, no.~9, pp. 577--591, 2007.

\bibitem{siow2022learning}
J.~K. Siow, S.~Liu, X.~Xie, G.~Meng, and Y.~Liu, ``Learning program semantics with code representations: An empirical study,'' in \emph{2022 IEEE international conference on software analysis, evolution and reengineering (SANER)}.\hskip 1em plus 0.5em minus 0.4em\relax IEEE, 2022, pp. 554--565.

\bibitem{wang2018ccaligner}
P.~Wang, J.~Svajlenko, Y.~Wu, Y.~Xu, and C.~K. Roy, ``Ccaligner: a token based large-gap clone detector,'' in \emph{Proceedings of the 40th International Conference on Software Engineering}, 2018, pp. 1066--1077.

\bibitem{golubev2021multi}
Y.~Golubev, V.~Poletansky, N.~Povarov, and T.~Bryksin, ``Multi-threshold token-based code clone detection,'' in \emph{2021 IEEE International Conference on Software Analysis, Evolution and Reengineering (SANER)}.\hskip 1em plus 0.5em minus 0.4em\relax IEEE, 2021, pp. 496--500.

\bibitem{wei2017supervised}
H.~Wei and M.~Li, ``Supervised deep features for software functional clone detection by exploiting lexical and syntactical information in source code.'' in \emph{IJCAI}, 2017, pp. 3034--3040.

\bibitem{zhang2019novel}
J.~Zhang, X.~Wang, H.~Zhang, H.~Sun, K.~Wang, and X.~Liu, ``A novel neural source code representation based on abstract syntax tree,'' in \emph{2019 IEEE/ACM 41st International Conference on Software Engineering (ICSE)}.\hskip 1em plus 0.5em minus 0.4em\relax IEEE, 2019, pp. 783--794.

\bibitem{jiang2022hierarchical}
Y.~Jiang, X.~Su, C.~Treude, and T.~Wang, ``Hierarchical semantic-aware neural code representation,'' \emph{Journal of Systems and Software}, vol. 191, p. 111355, 2022.

\bibitem{liu2023learning}
J.~Liu, J.~Zeng, X.~Wang, and Z.~Liang, ``Learning graph-based code representations for source-level functional similarity detection,'' in \emph{2023 IEEE/ACM 45th International Conference on Software Engineering (ICSE)}.\hskip 1em plus 0.5em minus 0.4em\relax IEEE, 2023, pp. 345--357.

\bibitem{wang2020detecting}
W.~Wang, G.~Li, B.~Ma, X.~Xia, and Z.~Jin, ``Detecting code clones with graph neural network and flow-augmented abstract syntax tree,'' in \emph{2020 IEEE 27th International Conference on Software Analysis, Evolution and Reengineering (SANER)}.\hskip 1em plus 0.5em minus 0.4em\relax IEEE, 2020, pp. 261--271.

\bibitem{mehrotra2023improving}
N.~Mehrotra, A.~Sharma, A.~Jindal, and R.~Purandare, ``Improving cross-language code clone detection via code representation learning and graph neural networks,'' \emph{IEEE Transactions on Software Engineering}, 2023.

\bibitem{zhang2025machine}
Z.~Zhang and T.~Saber, ``Machine learning approaches to code similarity measurement: A systematic review,'' \emph{IEEE Access}, 2025.

\bibitem{yuan2020local}
Y.~Yuan, W.~Kong, G.~Hou, Y.~Hu, M.~Watanabe, and A.~Fukuda, ``From local to global semantic clone detection,'' in \emph{2019 6th International Conference on Dependable Systems and Their Applications (DSA)}.\hskip 1em plus 0.5em minus 0.4em\relax IEEE, 2020, pp. 13--24.

\bibitem{yu2023graph}
D.~Yu, Q.~Yang, X.~Chen, J.~Chen, and Y.~Xu, ``Graph-based code semantics learning for efficient semantic code clone detection,'' \emph{Information and Software Technology}, vol. 156, p. 107130, 2023.

\bibitem{li2020semantic}
B.~Li, C.~Ye, S.~Guan, and H.~Zhou, ``Semantic code clone detection via event embedding tree and gat network,'' in \emph{2020 IEEE 20th International Conference on Software Quality, Reliability and Security (QRS)}.\hskip 1em plus 0.5em minus 0.4em\relax IEEE, 2020, pp. 382--393.

\bibitem{zhang2024fsd}
L.~Zhang, S.~Luo, L.~Pan, Z.~Wu, and K.~Gong, ``Fsd-clcd: Functional semantic distillation graph learning for cross-language code clone detection,'' \emph{Engineering Applications of Artificial Intelligence}, vol. 133, p. 108199, 2024.

\bibitem{zhao2022precise}
Z.~Zhao, B.~Yang, G.~Li, H.~Liu, and Z.~Jin, ``Precise learning of source code contextual semantics via hierarchical dependence structure and graph attention networks,'' \emph{Journal of Systems and Software}, vol. 184, p. 111108, 2022.

\bibitem{yuan2022java}
D.~Yuan, S.~Fang, T.~Zhang, Z.~Xu, and X.~Luo, ``Java code clone detection by exploiting semantic and syntax information from intermediate code-based graph,'' \emph{IEEE Transactions on Reliability}, vol.~72, no.~2, pp. 511--526, 2022.

\bibitem{zhang2025ast}
Z.~Zhang and T.~Saber, ``Ast-enhanced or ast-overloaded? the surprising impact of hybrid graph representations on code clone detection,'' \emph{arXiv preprint arXiv:2506.14470}, 2025.

\bibitem{svajlenko2016bigcloneeval}
J.~Svajlenko and C.~K. Roy, ``Bigcloneeval: A clone detection tool evaluation framework with bigclonebench,'' in \emph{2016 IEEE international conference on software maintenance and evolution (ICSME)}.\hskip 1em plus 0.5em minus 0.4em\relax IEEE, 2016, pp. 596--600.

\bibitem{vaswani2017attention}
A.~Vaswani, N.~Shazeer, N.~Parmar, J.~Uszkoreit, L.~Jones, A.~N. Gomez, {\L}.~Kaiser, and I.~Polosukhin, ``Attention is all you need,'' \emph{Advances in neural information processing systems}, vol.~30, 2017.

\bibitem{velivckovic2017graph}
P.~Veli{\v{c}}kovi{\'c}, G.~Cucurull, A.~Casanova, A.~Romero, P.~Lio, and Y.~Bengio, ``Graph attention networks,'' \emph{arXiv preprint arXiv:1710.10903}, 2017.

\bibitem{dosovitskiy2020image}
A.~Dosovitskiy, L.~Beyer, A.~Kolesnikov, D.~Weissenborn, X.~Zhai, T.~Unterthiner, M.~Dehghani, M.~Minderer, G.~Heigold, S.~Gelly \emph{et~al.}, ``An image is worth 16x16 words: Transformers for image recognition at scale,'' \emph{arXiv preprint arXiv:2010.11929}, 2020.

\bibitem{lu2021code}
Z.~Lu, R.~Li, H.~Hu, and W.-a. Zhou, ``A code clone detection algorithm based on graph convolution network with ast tree edge,'' in \emph{2021 IEEE 21st International Conference on Software Quality, Reliability and Security Companion (QRS-C)}.\hskip 1em plus 0.5em minus 0.4em\relax IEEE, 2021, pp. 1027--1032.

\bibitem{mehrotra2021modeling}
N.~Mehrotra, N.~Agarwal, P.~Gupta, S.~Anand, D.~Lo, and R.~Purandare, ``Modeling functional similarity in source code with graph-based siamese networks,'' \emph{IEEE Transactions on Software Engineering}, vol.~48, no.~10, pp. 3771--3789, 2021.

\bibitem{zhang2023efficient}
A.~Zhang, L.~Fang, C.~Ge, P.~Li, and Z.~Liu, ``Efficient transformer with code token learner for code clone detection,'' \emph{Journal of Systems and Software}, vol. 197, p. 111557, 2023.

\bibitem{fang2020functional}
C.~Fang, Z.~Liu, Y.~Shi, J.~Huang, and Q.~Shi, ``Functional code clone detection with syntax and semantics fusion learning,'' in \emph{Proceedings of the 29th ACM SIGSOFT international symposium on software testing and analysis}, 2020, pp. 516--527.

\bibitem{xu2021sccd}
K.~Xu and Y.~Liu, ``Sccd-gan: An enhanced semantic code clone detection model using gan,'' in \emph{2021 IEEE 4th International Conference on Electronics and Communication Engineering (ICECE)}.\hskip 1em plus 0.5em minus 0.4em\relax IEEE, 2021, pp. 16--22.

\bibitem{swilam2023cross}
Z.~Swilam, A.~Hamdy, and A.~Pester, ``Cross-language code clone detection using abstract syntax tree and graph neural network,'' in \emph{2023 International Conference on Computer and Applications (ICCA)}.\hskip 1em plus 0.5em minus 0.4em\relax IEEE, 2023, pp. 1--5.

\bibitem{hua2020fcca}
W.~Hua, Y.~Sui, Y.~Wan, G.~Liu, and G.~Xu, ``Fcca: Hybrid code representation for functional clone detection using attention networks,'' \emph{IEEE Transactions on Reliability}, vol.~70, no.~1, pp. 304--318, 2020.

\bibitem{baxter1998clone}
I.~D. Baxter, A.~Yahin, L.~Moura, M.~Sant'Anna, and L.~Bier, ``Clone detection using abstract syntax trees,'' in \emph{Proceedings. International Conference on Software Maintenance (Cat. No. 98CB36272)}.\hskip 1em plus 0.5em minus 0.4em\relax IEEE, 1998, pp. 368--377.

\bibitem{javalang}
D.~Hovemeyer, ``javalang: Pure python library for parsing java source code,'' \url{https://github.com/c2nes/javalang}, 2015, accessed: 2025-07-21.

\bibitem{koppel2022automatically}
J.~Koppel, J.~Kearl, and A.~Solar-Lezama, ``Automatically deriving control-flow graph generators from operational semantics,'' \emph{Proceedings of the ACM on Programming Languages}, vol.~6, no. ICFP, pp. 742--771, 2022.

\bibitem{li2018deeper}
Q.~Li, Z.~Han, and X.-M. Wu, ``Deeper insights into graph convolutional networks for semi-supervised learning,'' in \emph{Proceedings of the AAAI conference on artificial intelligence}, vol.~32, no.~1, 2018.

\bibitem{bresson2017residual}
X.~Bresson and T.~Laurent, ``Residual gated graph convnets,'' \emph{arXiv preprint arXiv:1711.07553}, 2017.

\bibitem{vinyals2015order}
O.~Vinyals, S.~Bengio, and M.~Kudlur, ``Order matters: Sequence to sequence for sets,'' \emph{arXiv preprint arXiv:1511.06391}, 2015.

\bibitem{svajlenko2014towards}
J.~Svajlenko, J.~F. Islam, I.~Keivanloo, C.~K. Roy, and M.~M. Mia, ``Towards a big data curated benchmark of inter-project code clones,'' in \emph{2014 IEEE international conference on software maintenance and evolution}.\hskip 1em plus 0.5em minus 0.4em\relax IEEE, 2014, pp. 476--480.

\bibitem{googlecodejam}
2016, google Code Jam https://code.google.com/codejam/contests.html.

\bibitem{zhao2018deepsim}
G.~Zhao and J.~Huang, ``Deepsim: deep learning code functional similarity,'' in \emph{Proceedings of the 2018 26th ACM joint meeting on european software engineering conference and symposium on the foundations of software engineering}, 2018, pp. 141--151.

\bibitem{jia2025code}
C.~Jia, Y.~Zhan, T.~Zhao, H.~Ye, and M.~Zhou, ``Code clone detection via an alphafold-inspired framework,'' \emph{arXiv preprint arXiv:2507.15226}, 2025.

\bibitem{paszke2019pytorch}
A.~Paszke, S.~Gross, F.~Massa, A.~Lerer, J.~Bradbury, G.~Chanan, T.~Killeen, Z.~Lin, N.~Gimelshein, L.~Antiga \emph{et~al.}, ``Pytorch: An imperative style, high-performance deep learning library,'' \emph{Advances in neural information processing systems}, vol.~32, 2019.

\bibitem{fey2019fast}
M.~Fey and J.~E. Lenssen, ``Fast graph representation learning with pytorch geometric,'' \emph{arXiv preprint arXiv:1903.02428}, 2019.

\bibitem{feng2020codebert}
Z.~Feng, D.~Guo, D.~Tang, N.~Duan, X.~Feng, M.~Gong, L.~Shou, B.~Qin, T.~Liu, D.~Jiang \emph{et~al.}, ``Codebert: A pre-trained model for programming and natural languages,'' \emph{arXiv preprint arXiv:2002.08155}, 2020.

\bibitem{yu2019neural}
H.~Yu, W.~Lam, L.~Chen, G.~Li, T.~Xie, and Q.~Wang, ``Neural detection of semantic code clones via tree-based convolution,'' in \emph{2019 IEEE/ACM 27th International Conference on Program Comprehension (ICPC)}.\hskip 1em plus 0.5em minus 0.4em\relax IEEE, 2019, pp. 70--80.

\bibitem{dou2024cc2vec}
S.~Dou, Y.~Wu, H.~Jia, Y.~Zhou, Y.~Liu, and Y.~Liu, ``Cc2vec: Combining typed tokens with contrastive learning for effective code clone detection,'' \emph{Proceedings of the ACM on Software Engineering}, vol.~1, no. FSE, pp. 1564--1584, 2024.

\bibitem{mou2016convolutional}
L.~Mou, G.~Li, L.~Zhang, T.~Wang, and Z.~Jin, ``Convolutional neural networks over tree structures for programming language processing,'' in \emph{Proceedings of the AAAI conference on artificial intelligence}, vol.~30, no.~1, 2016.

\end{thebibliography}

\begin{IEEEbiography}[{\includegraphics[width=1in,height=1.25in,clip,keepaspectratio]{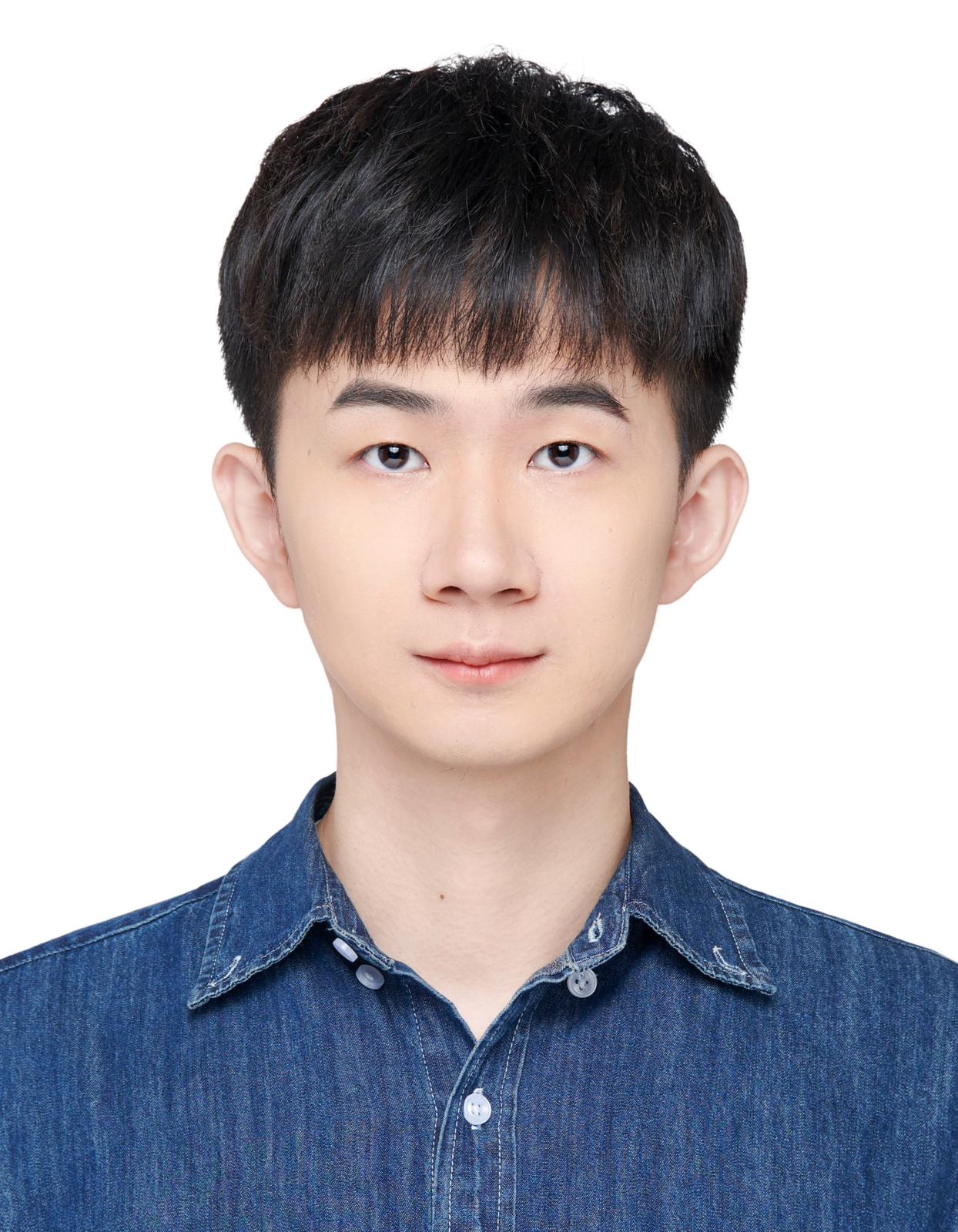}}]{Zixian Zhang} is a PhD student in the SFI Centre for Research Training in Artificial Intelligence (CRT-AI) the School of Computer Science at University of Galway, Ireland, under the supervision of Dr Takfarinas Saber. His main research topic is on the design and application of novel machine learning techniques for the assessment and measurement of code similarity. 
\end{IEEEbiography}

\begin{IEEEbiography}[{\includegraphics[width=1in,height=1.25in,clip,keepaspectratio]{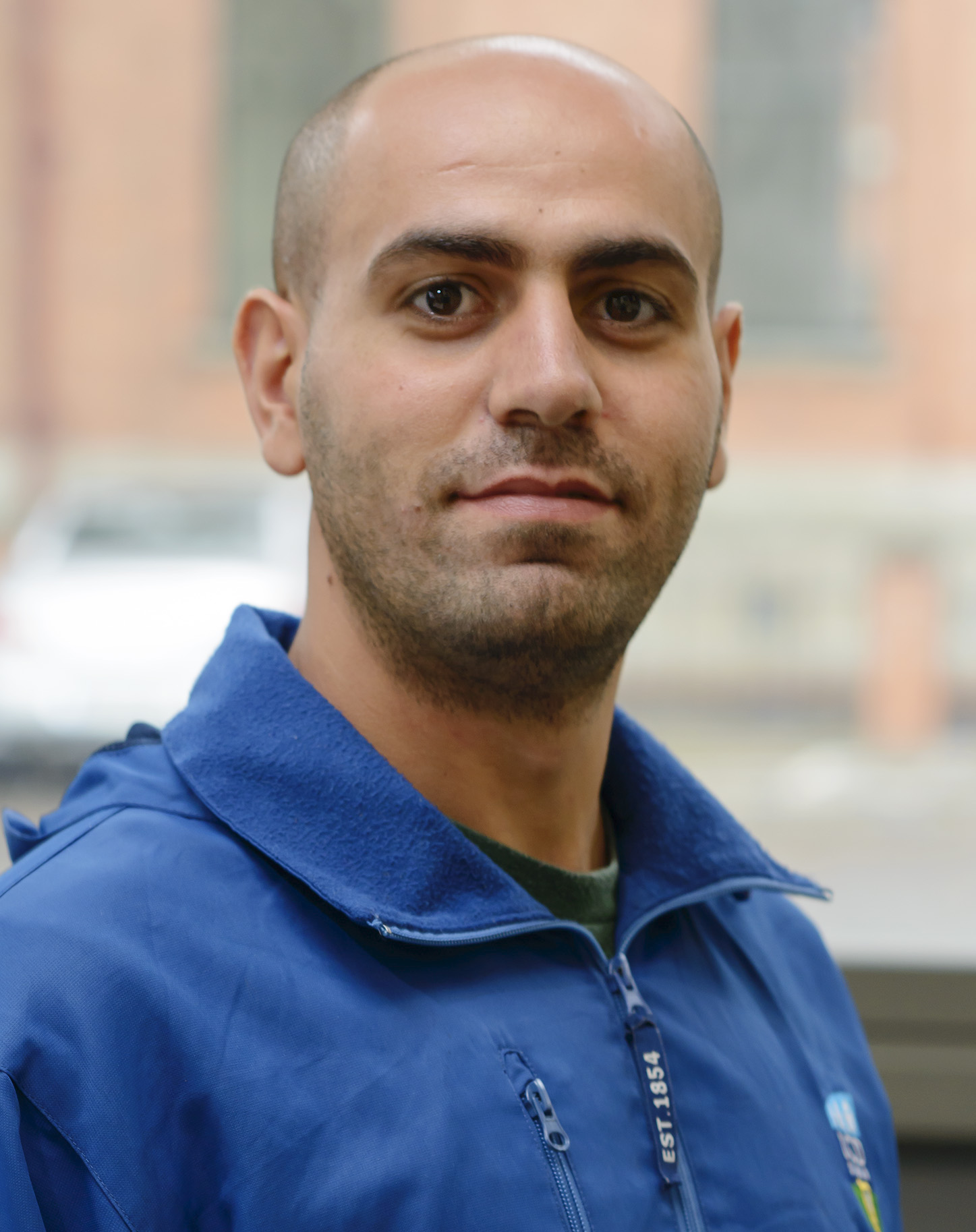}}]{Takfarinas Saber} holds a PhD (2017) in Computer Science from University College Dublin, Ireland. He is currently a Lecturer in Computer Science at the University of Galway, Ireland. His area of expertise is in the optimisation of Complex Software Systems. He designs and applies novel Artificial Intelligence techniques from Operations Research, Machine Learning, and Evolutionary Computation/Learning to advance the Engineering and Testing of software systems such as Smart Cities, Distributed Systems, and Wireless Communication Networks.
\end{IEEEbiography}

\end{document}